\documentclass[runningheads]{svmult}

\usepackage{makeidx}   
\usepackage{graphicx}  
\input{psfig.sty}
\usepackage{subeqnar}  
\usepackage{multicol}  
\usepackage{physprbb}  
\makeindex             

\def\spose#1{\hbox to 0pt{#1\hss}}
\def\lta{\mathrel{\spose{\lower 3pt\hbox{$\mathchar"218$}} \raise 2.0pt\hbox{$\mathchar"13C$}}}
\def\gta{\mathrel{\spose{\lower 3pt\hbox{$\mathchar"218$}} \raise 2.0pt\hbox{$\mathchar"13E$}}}
\newcommand{\etal}{et~al.\ }
\def\msun{\,{\rm M_\odot}}
\makeatletter

\newenvironment{figurehere}
  {\def\@captype{figure}}
  {}
\makeatother
\newcommand\beq{\begin{equation}}
\newcommand\eeq{\end{equation}}
\def\HII{\hbox{H~$\scriptstyle\rm II\ $}}

\begin{document}
\title*{From the earliest seeds to today's supermassive black holes} 
\toctitle{Earliest seed MBHs} 
%
%
\titlerunning{Earliest seed MBHs}
%
\author{Piero Madau}
\authorrunning{Piero Madau}
\institute{Department of Astronomy and Astrophysics, University of California,
Santa Cruz CA 95064, USA}

\maketitle              

\begin{abstract}
I review scenarios for the assembly of supermassive black holes (MBHs) at the center 
of galaxies that trace their hierarchical build-up far up in the dark matter halo 
``merger tree". Monte Carlo realizations of the merger hierarchy in a $\Lambda$CDM 
cosmology, coupled to semi-analytical recipes, are a powerful tool to follow the merger 
history of halos and the dynamics and growth of the MBHs they host. X-ray photons from 
miniquasars powered by intermediate-mass ``seed" holes may permeate the universe more 
uniformly than EUV radiation, make the low-density diffuse intergalactic medium warm 
and weakly ionized prior to the epoch of reionization breakthrough, and set an entropy floor.
The spin distribution of MBHs is determined by gas accretion, and is predicted to be heavily 
skewed towards fast-rotating 
Kerr holes, already in place at early epochs, and not to change significantly below 
redshift 5. Decaying MBH binaries may shape the innermost central regions of galaxies
and should be detected in significant numbers by {\it LISA}. 
\end{abstract}

\section{Massive black holes and galaxy formation}

The strong link observed between the masses of supermassive black holes (MBHs) 
at the center of most galaxies and the gravitational potential wells
that host them suggests a fundamental mechanism for assembling black holes and
forming spheroids in galaxy halos. The $m_{\rm BH}$-$\sigma$ relation \cite{fm}
\cite{g00} implies a rough proportionality between MBH mass and the mass of the
baryonic component of the bulge. It is not yet understood whether this relation was set
in primordial structures, and consequently how it is maintained throughout cosmic
time with such a small dispersion, or indeed which physical processes established
such a correlation in the first place \cite{sr}\cite{hk}\cite{bs}.

In cosmologies dominated by cold dark matter (CDM) galaxy halos experience multiple
mergers during their lifetime, with those between comparable-mass systems
(``major mergers'') expected to result in the formation of elliptical galaxies
\cite{h92}. Simple models in which MBHs are also assumed to grow during 
major mergers and to be present in every galaxy at any
redshift -- while only a fraction of them is ``active'' at any given time -- have
been shown to explain many aspects of the observed evolution of quasars 
\cite{chr}\cite{cv}\cite{kh}. The coevolution of MBHs and their host galaxies
in hierarchical structure formation scenarios gives origin to a number of 
of important questions, most notably:

$\bullet$ {\it Did the first MBHs form in subgalactic units far up in the
merger hierarchy, well before the bulk of the stars observed today?}
The seeds of the $z\sim 6$ quasars discovered in the {\it Sloan Digital
Sky Survey} had to appear at very high redshift,
$z\gta 10$, if they are accreting no faster than the Eddington rate.
In hierarchical cosmologies, the ubiquity of MBHs in nearby luminous 
galaxies can arise even if only a small fraction of halos harbor MBHs 
at very high redshift \cite{mhn}. 

$\bullet$ {\it How massive were the initial seeds, and is there a population of 
relic pregalactic MBHs lurking in present-day galaxy halos?} A clue to these
questions may lie in the numerous population of ultraluminous off-nuclear 
(``non-AGN'') X-ray sources that have been detected in nearby galaxies \cite{mc}. 
Assuming isotropic emission, the inferred masses of these ``ULXs" may suggest 
intermediate-mass black holes with masses $\gta$ a few hundred $\msun$ 
\cite{cm}\cite{k01}.

$\bullet$ {\it Can coalescing MBH binaries at very high redshift be detected 
in significant 
numbers by the planned {\it Laser Interferometer Space Antenna} ({\it LISA})?}
If MBHs were common in the past (as implied by the notion that many distant galaxies
harbor active nuclei for a short period of their life), and if their host
galaxies undergo multiple mergers, then MBH binaries
will inevitably form in large numbers during cosmic history. MBH pairs that
are able to coalesce in less than a Hubble time will give origin to the
loudest gravitational wave events in the universe.

$\bullet$ If was first proposed by \cite{emo} that the heating
of the surrounding stars by a decaying MBH pair would create a low-density
core out of a preexisting cuspy (e.g. $\rho_*\propto r^{-2}$) stellar density profile.
If stellar dynamical processes can efficiently drive wide MBH binaries 
to the gravitational wave (GW) emission stage, {\it what is the cumulative dynamical 
effect of multiple black hole mergers on galaxy stellar cusps?} 

$\bullet$ Active galactic nuclei powered by supermassive holes keep the universe 
ionized
at $z\lta 4$, structure the intergalactic medium (IGM), and probably regulate star 
formation in their host galaxies. Intermediate-mass holes accreting gas 
from the surrounding medium may shine as ``miniquasars" at redshifts as high as 
$z\sim 20$. {\it What is the thermodynamic effect of miniquasars on the IGM 
at early times?}

$\bullet$ Besides their masses, astrophysical black holes are completely characterized 
by their spins, $S=aGm_{\rm BH}/c$, $0\le a/m_{\rm BH}\le 1$. The spin of a MBH is 
expected to have a 
significant effect on its observational manifestation, such as the efficiency of
converting accreted mass into radiation and the existence and direction of jets in active nuclei.
{\it What is the expected distribution of MBH spins and how does this evolve with 
cosmic time?}

In this talk I will review some recent developments in our understanding of the assembly, 
growth, emission history, and environmental impact of MBHs from early epochs to 
the present. Unless otherwise stated, 
all the results shown below refer to the currently favoured (by a variety of 
observations) $\Lambda$CDM world model with $\Omega_M=0.3$,
$\Omega_\Lambda=0.7$, $h=0.7$, $\Omega_b=0.045$, $\sigma_8=0.93$, and $n=1$.

\section{MBHs as Population III remnants}

The first stars in the universe must have formed out of metal-free gas, in dark
matter ``minihalos''  of total mass $\gta 5\times 10^5\msun$ \cite{fc} 
condensing from the high-$\sigma$ peaks of the primordial 
density field at redshift $z=20-30$. Numerical
simulations of the fragmentation of primordial clouds in standard CDM theories
all show the formation of Jeans unstable clumps with masses exceeding a few hundred
solar masses; because of the slow subsonic contraction -- a regime set up by the
main gas coolant, molecular hydrogen -- further fragmentation into
sub-components is not seen, and a single very massive star forms from the 
inside out \cite{bcl}\cite{abn}\cite{op}. 

\begin{figurehere}
\vspace{0.4cm}
\centerline{
\psfig{figure=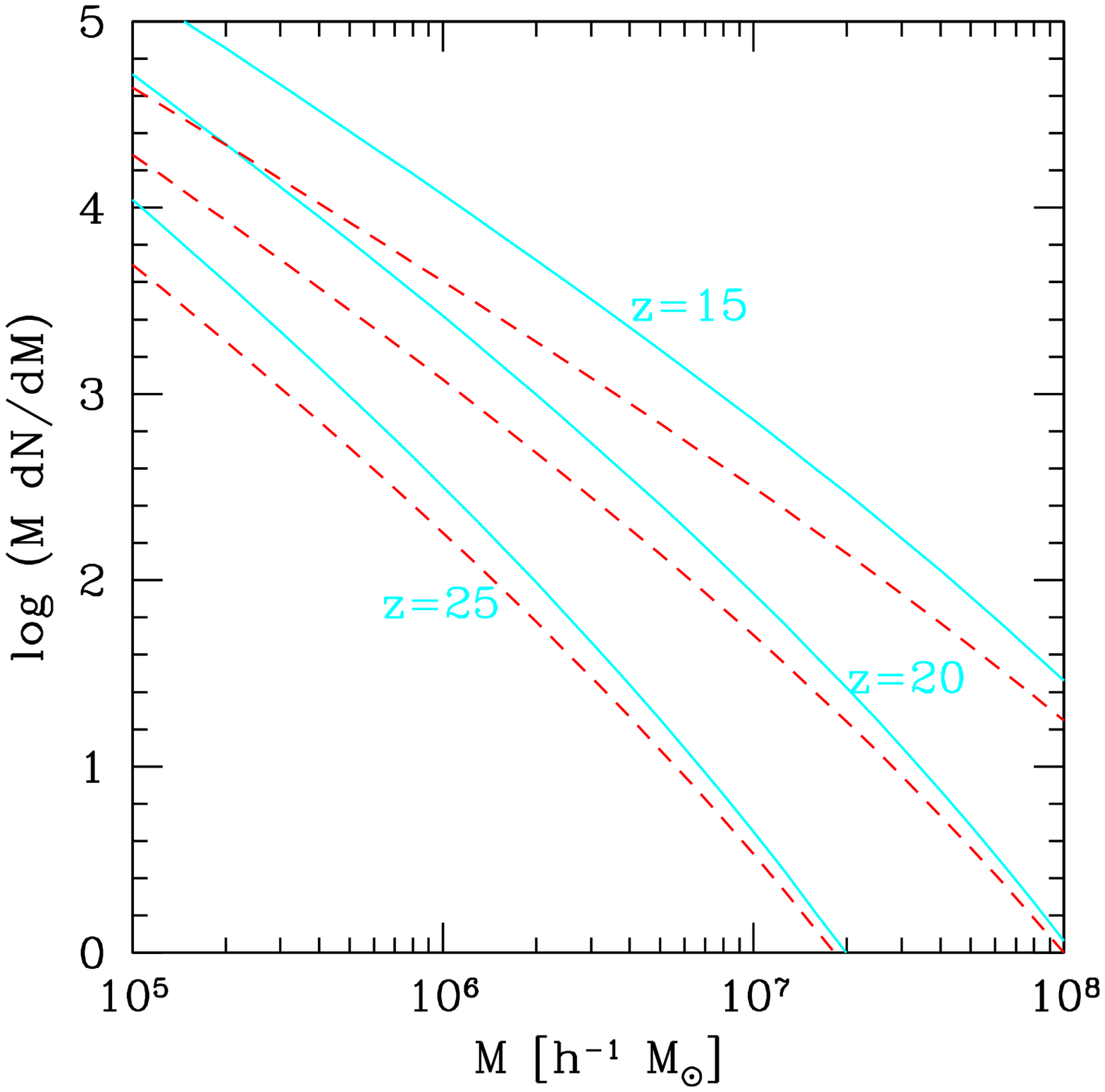,width=3.0in}}
\vspace{0.0cm}
\caption{\footnotesize
Mass function of minihalos of mass $M$ formed at
$z=15, 20, 25$ which, by the later time $z_0$, will have merged into
a more massive halo of total mass $M_0$. {\it Solid curves:} $z_0=0.8$,
$M_0=10^{12}\,h^{-1}\,\msun$ (``Milky Way'' halo). {\it Dashed curves:}
$z_0=3.5$, $M_0=2\times 10^{11}\,h^{-1}\,\msun$ (older ``bulge''). 
If only one seed hole formed in each $\sim 10^6\,\msun$ minihalo collapsing at 
$z\sim 20$ (and triple hole interactions and binary coalescences were 
neglected), several thousands relic IMBHs and their descendants would
be orbiting throughout present-day galaxy halos \cite{mr}\cite{its}.
}
\label{fig1}
\vspace{0.4cm}
\end{figurehere}

At zero metallicity mass loss through radiatively-driven stellar
winds or nuclear-powered stellar pulsations is expected to be negligible, and 
Population III stars will likely die losing only a small fraction of their mass
(except for $100<m_*<140\,\msun$). Nonrotating
very massive stars in the mass window $140\lta m_*\lta 260\,\msun$ will disappear
as pair-instability supernovae \cite{bac}, leaving no compact remnants and
polluting the universe with the first heavy elements \cite{sch}\cite{oh01}.
Stars with $40<m_*<140\,\msun$ and $m_*>260\,\msun$ are predicted instead
to collapse to black holes with masses exceeding half of the initial
stellar mass \cite{hw}. Barring any fine tuning of the initial mass function
of Pop III stars, 
intermediate-mass black holes (IMBHs) -- with masses above the 4--18$\,\msun$ range of 
known ``stellar-mass" holes  -- may then be the inevitable endproduct of the first
episodes of pregalactic star formation \cite{mr}.
Since they form in high-$\sigma$ rare density peaks, relic IMBHs are expected
to cluster in the bulges of
present-day galaxies as they become incorporated through a series of
mergers into larger and larger systems (see Fig. \ref{fig1}). 
The presence of a small cluster of IMBHs in galaxy nuclei may have several interesting 
consequences associated with tidal captures of ordinary
stars (likely followed by disruption), capture by the central supermassive hole, 
and gravitational wave radiation from such coalescences \cite{m04}.
Accreting pregalactic IMBHs may be detectable as ultra-luminous, off-nuclear
X-ray sources \cite{mr}\cite{k04}.

\section{The first miniquasars}

Physical conditions in the central potential wells of young and gas-rich 
protogalaxies may have been propitious for black hole gas accretion.
Perhaps seed black holes grew efficiently in small minihalos just above
the cosmological Jeans mass (with shallow potential wells), or maybe gas
accretion had to await the buildup of more massive galaxies 
(with virial temperatures above the threshold for atomic cooling). 
This issue is important for the detectability of high-$z$ miniquasars: it also 
determines whether the radiation background at very high redshifts had an X-ray component
component able to preheat and partially ionize the IGM.

As mentioned above, gas condensation in the first
baryonic objects is possible through the formation of H$_2$ molecules,
which cool via roto-vibrational transitions down to temperatures of a few 
hundred kelvins. In the absence of a UV
photodissociating flux and of ionizing X-ray radiation, three-dimensional
simulations of early structure formation show that the fraction of
cold, dense gas available for accretion onto seed holes or star formation
exceeds 20\% for halos more massive than $10^6\,\msun$ \cite{mba}. 
On th eother hand, a zero-metallicity  progenitor star in the range $40<m_*<500\,\msun$ 
emits about 70,000 
photons above 1 ryd per stellar baryon \cite{scha}. The ensuing ionization
front will completely overrun the host halo, photoevaporating most of the 
surrounding gas \cite{wan}. Black hole remnants of the first stars that
created \HII regions are then unlikely to accrete significant mass until new 
cold material will be made available through the hierarchical merging of 
many gaseous subunits.

Accretion onto IMBHs may be an attractive way to (partially) reionize the low-density
IGM \cite{mad04}\cite{ric}. A large fraction of the UV radiation from massive
stars may not escape the dense sites of star formation, or may be deposited
locally in halo gas that recombines almost immediately. The harder radiation
emitted from miniquasars is instead more likely to escape from the hosts
into intergalactic space, and may then produce more `durable' (albeit
partial) ionization in the diffuse IGM. High-resolution hydrodynamics simulations 
of early structure formation in $\Lambda$CDM cosmologies are a powerful tool to
track in detail the thermal and ionization history of a clumpy IGM and guide studies 
of early reheating. We \cite{km} have used {\sl Enzo}, an adaptive mesh refinement (AMR), 
grid-based hybrid (hydro$+$N-body) code developed by Bryan \& Norman 
(see http://cosmos.ucsd.edu/enzo/) to solve the cosmological hydrodynamics equations
and simulate the effect of a miniquasar turning on at very high redshift in a volume 1 Mpc
on a side (comoving). We first identify in a low-resolution pure N-body simulation 
the Lagrangian volume of a resolved protogalactic halo with a total mass 
$7\times 10^5\,\msun$ at $z=25$, above the cosmological Jeans mass. We then generate
new initial conditions with an 128$^3$ initial static grid that covers a 0.5 Mpc
volume centered around the identified high-$\sigma$ peak. During the evolution,
refined grids (for a maximum of 5 additional levels) are introduced with twice 
the spatial resolution of the parent 
(coarser) grid to home in, with progressively finer resolution, on the densest parts
of the ``cosmic web''. The simulation follows the non-equilibrium chemistry of the
dominant nine species (H, H$^+$, H$^-$, e, He, He$^+$, He$^{++}$, H$_2$, and H$_2^+$) 
in primordial gas, and includes radiative losses from atomic and molecular line 
cooling. 

\begin{figurehere}
\vspace{0.4cm}
\centerline{
\psfig{figure=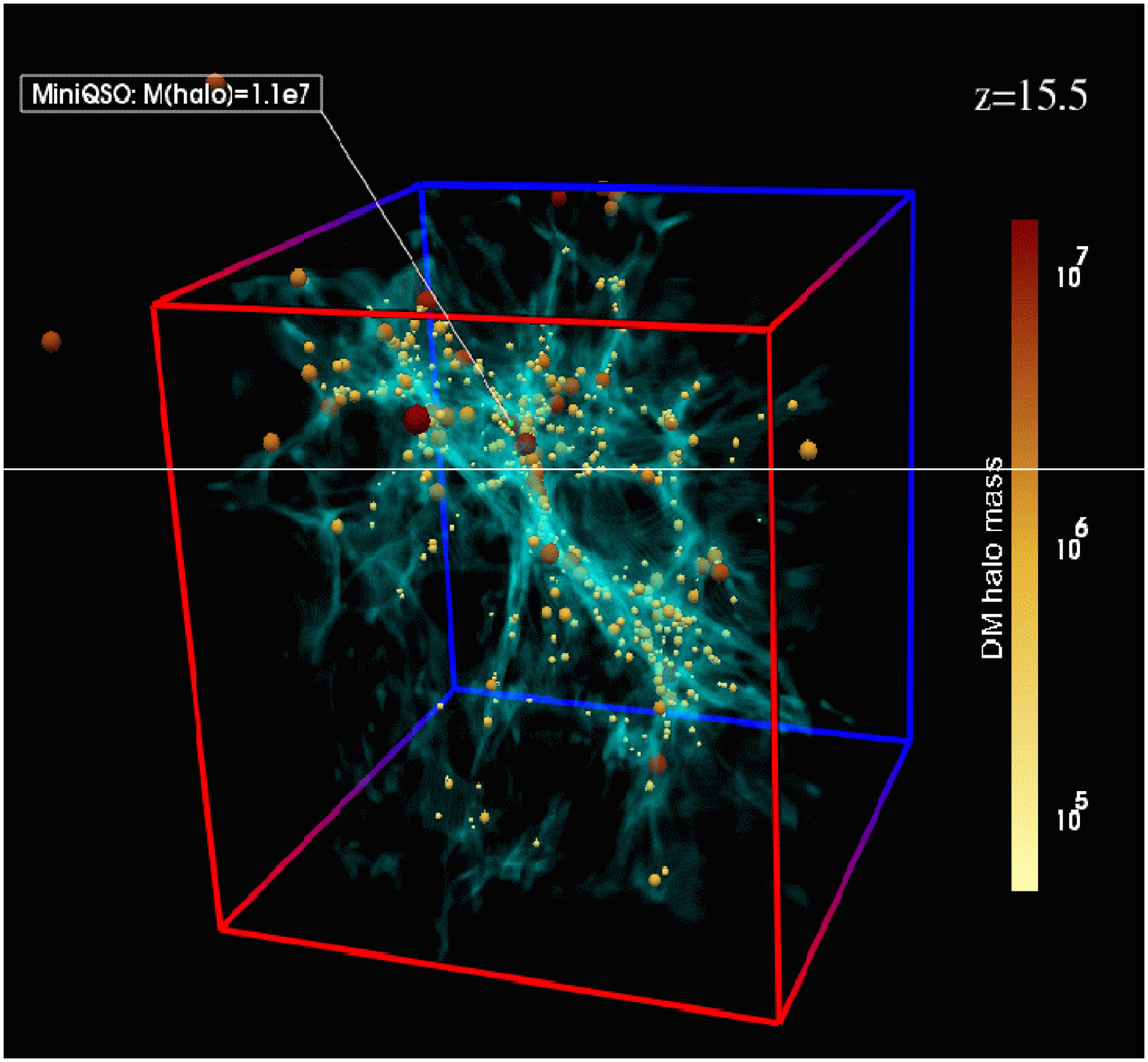,width=3.0in}}
\vspace{0.0cm}
\caption{\footnotesize Gas distribution (at overdensity 2) in the inner 0.5 Mpc 
of the simulation box at $z=15.5$. Dark matter halos of different masses 
(identified with a halo-finder algorithm) are plotted as dark spheres. 
}
\label{fig2}
\vspace{0.4cm}
\end{figurehere}

At $z=21$, a miniquasar powered by a 150 $\msun$ black hole accreting
at the Eddington rate is turned on in the protogalactic halo. The miniquasar 
shines for a few Salpeter times (i.e. down to $z\sim 15$) and is a copious source 
of soft X-ray photons, 
which permeate the IGM more uniformly than possible with extreme ultraviolet (EUV,
$\ge 13.6\,$eV) radiation \cite{oh} and make it warm and
weakly ionized prior to the epoch of reionization breakthrough \cite{vgs}.
A spectrum with $\nu\,L_\nu=$ const (like the nonthermal component observed
in ULXs) was assumed for photons with energies in the range 0.2-10 keV, to which 
the simulation box is transparent. X-rays alone do not produce a fully 
ionized medium, but partially photoionize the gas by repeated secondary ionizations.
A primary nonthermal photoelectron of energy $E=1\,$keV in
a medium with residual ionization (from the recombination epoch)
$x=2\times 10^{-4}$ will create over two dozens secondary electrons,
depositing a fraction $f_{\rm ion}\approx 37\%$ of its initial energy as
secondary ionizations of hydrogen, and only $f_{\rm heat}\approx 13\%$ as
heat \cite{svs}. 
The timescale for electron-electron encounters resulting
in a fractional energy loss $f=\Delta E/E$,
\beq
t_{\rm ee}\approx 140\,{\rm yr}~Ef\,\left({1+z\over 20}\right)^{-3}\,
\left({\ln\Lambda\over 20}\right)^{-1}\,x^{-1}
\eeq
(where $E$ is measured in keV), is typically much shorter that the
electron Compton
cooling timescale off cosmic microwave background (CMB) photons,
$t_C=(7\times 10^6\, {\rm yr})\,[(1+z)/20]^{-4}$,
and thus the primary photoelectron will
ionize and heat the surrounding medium before it is cooled by the CMB.
Once the IGM ionized fraction increases to $x\approx 0.1$,
the number of secondary ionizations per ionizing photon drops to a
few, and the bulk of the primary's energy goes
into heat ($f_{\rm heat}\approx 0.6$) via elastic Coulomb collisions
with thermal electrons.

\begin{figurehere}
\vspace{0.4cm}
\centerline{
\psfig{figure=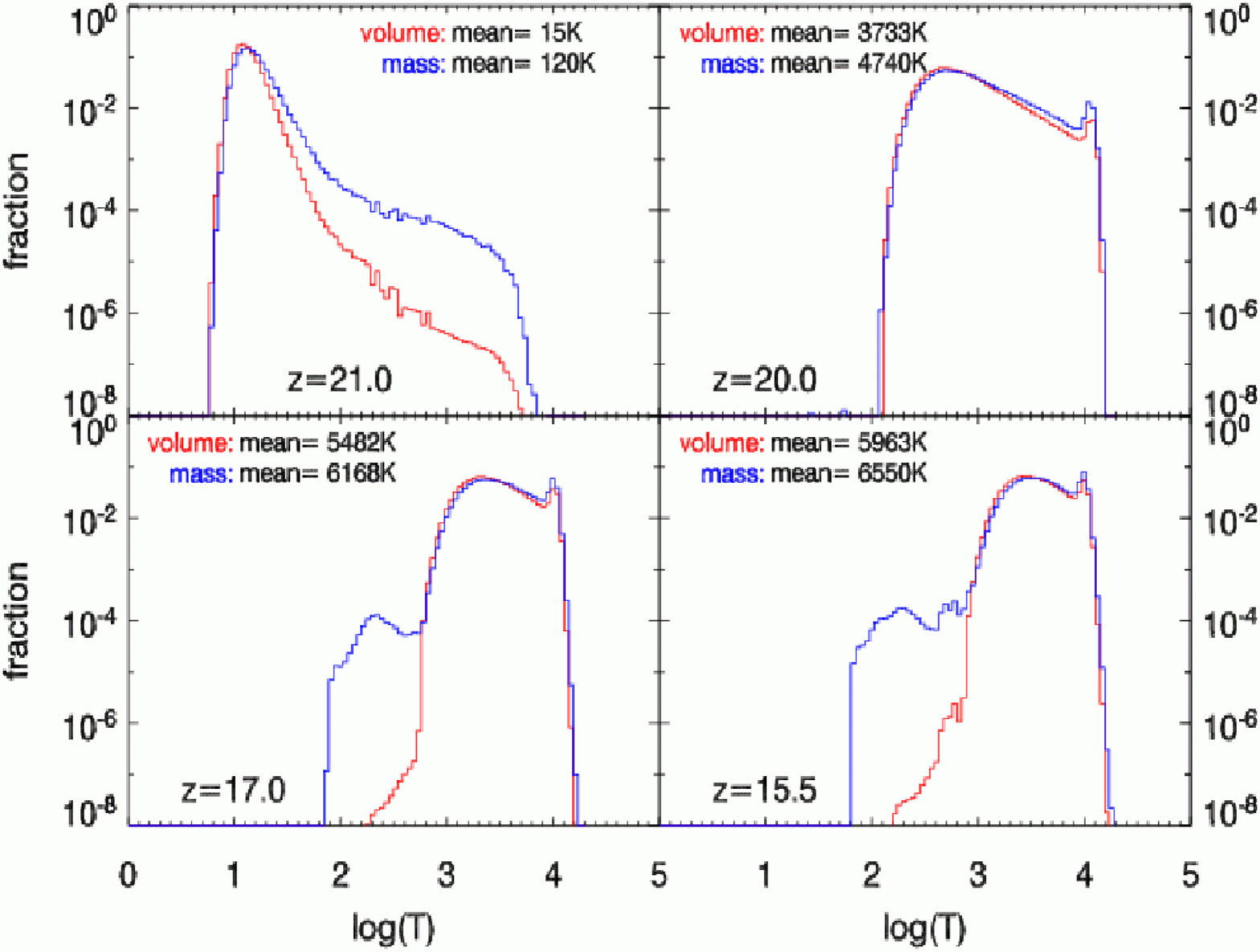,width=3.2in}}
\vspace{0.0cm}
\caption{\footnotesize Mass (upper curves) and volume (lower curves) weighted 
]gas fraction vs. gas temperature
in the simulation box. A miniquasar was turned on at $z=21$.  
}
\label{fig3}
\vspace{0.4cm}
\end{figurehere}

Figure \ref{fig3} shows the mass and volume-weighted gas fraction in the
simulation box vs. temperature as four different redshifts. The heating effect of X-rays from
the miniquasar is clearly seen, with gas temperatures between $10^3$ and $10^4$ K at $z<20$.
Strong Jeans mass filtering takes place, and subsequent minihalos will no longer be able to 
accrete gas due to the smoothing effect of gas pressure. The low-density IGM acquires 
a uniform ``entropy floor'' \cite{ohha} that (a) greatly reduces gas clumping, curtailing the
number of photons needed to maintain reionization, and (b) results
in significantly lower gas densities in the cores of minihalos that
suppress rapid H$_2$ formation. The latter effect may imply that X-rays inhibit rather than
enhance star formation \cite{har}\cite{mba}.

\section{MBH binaries and galaxy cores}

Frequent galaxy mergers will inevitably lead to the formation of MBH binaries.
As dark matter halos assemble, MBHs get incorporated into larger and larger 
halos, sink to the center owing to dynamical friction, accrete a fraction of the gas 
in the merger remnant to become supermassive, form a binary system, and 
eventually coalesce \cite{bbr}.
In a stellar background a ``hard" binary shrinks by capturing the stars that pass
close to the holes and ejecting them at much higher velocities, a super-elastic
scattering process that depletes the
nuclear region and turns a stellar ``cusp'' into a low-density core. Rapid 
coalescence eventually ensues due to the emission of gravitational radiation.
Observationally, there is clear evidence in early-type galaxies for a
systematic trend in the
distribution of surface brightness profiles, with faint ellipticals showing
steep power-law profiles (cusps), while bright ellipticals have much
shallower stellar cores \cite{faber}. Detailed N-body simulations have confirmed the
cusp-disruption effect of a hardening MBH binary \cite{milos}, but have shed
little light on why bright ellipticals have lower central concentrations
than faint ellipticals.

\begin{figurehere}
\vspace{0.4cm}
\centerline{
\psfig{file=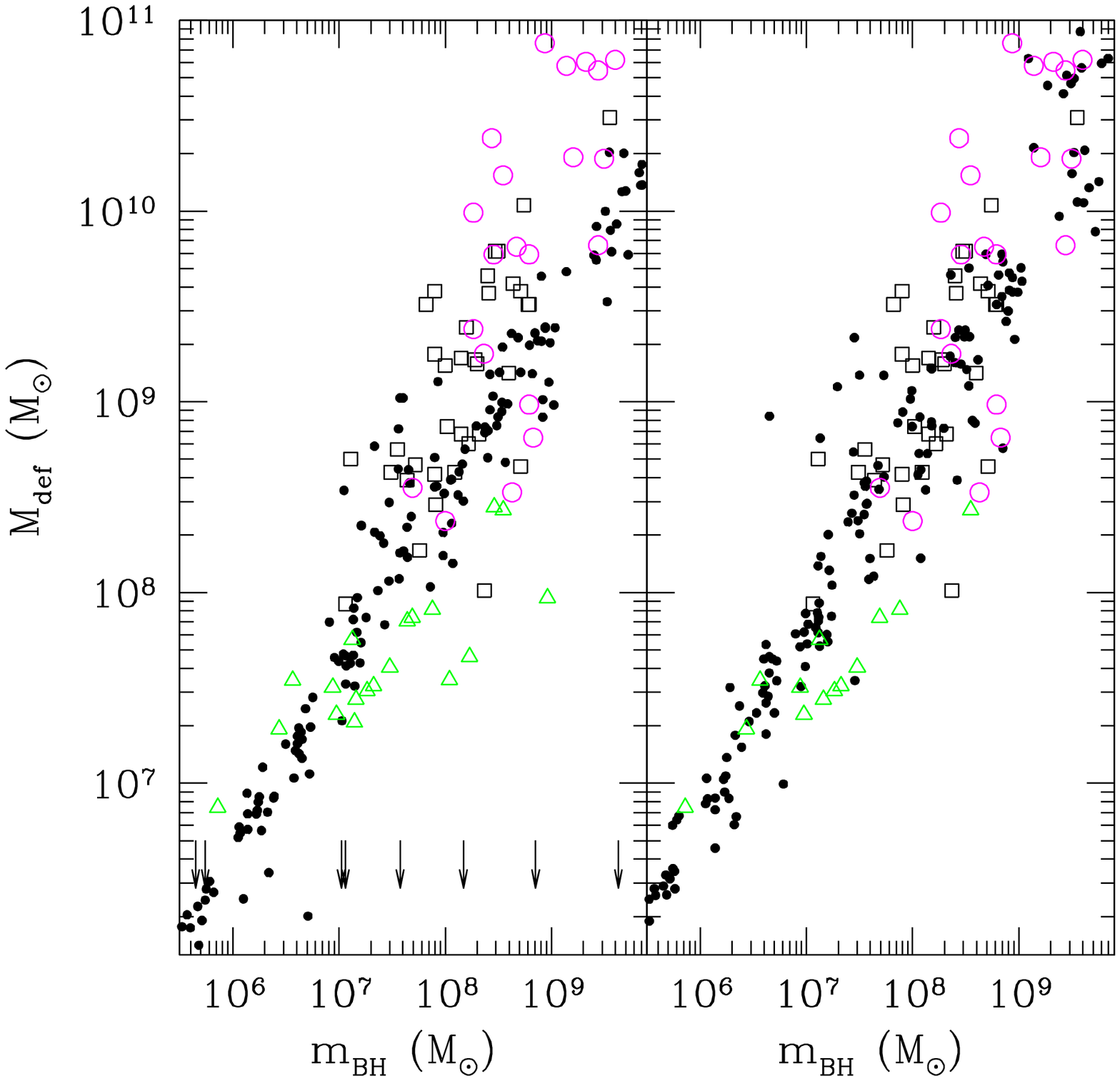,width=2.7in}}
\vspace{0.0cm}
\caption{\footnotesize Mass deficit produced at $z=0$ by shrinking MBHs
in our merger tree as a function of nuclear MBH mass ({\it filled dots}).
{\it Left panel:} cusp regeneration case. {\it Right panel:} core preservation case.
Galaxies without a core (i.e. those that have never
experienced a MBH-MBH merger or with their cusp recently regenerated) are
shown as vertical arrows at an arbitrary mass deficit of $10^6\msun$.
{\it Empty squares}: mass deficit inferred in a sample of galaxies by
\cite{milos02}. {\it Empty circles}: same for the ``core" galaxies
of \cite{faber}. {\it Empty triangles}: same for the
``cuspy" galaxies of \cite{faber}, assuming a flat core within
the upper limit on the core size. (From \cite{vmh}.)
} \label{fig4}
\end{figurehere}
\vspace{0.4cm}

The role of MBH binaries in shaping the central structure of galaxies is best
understood within the framework of a detailed model for the hierarchical
assembly of MBHs over cosmic history \cite{vhm}. Stellar cusps can be efficiently 
destroyed over cosmic time by decaying binaries if stellar dynamical processes are able
to shrink the binary down to a separation $\lta 10\%$ of the separation
at which the binary becomes hard. More massive halos have more massive
nuclear holes and experience more merging events than less massive galaxies:
hence they suffer more from the eroding action of binary MBHs and have
larger cores. In \cite{vmh} we found that a model in which the
effect of the hierarchy
of MBH interactions is cumulative and cores are preserved during galaxy
mergers produces at the present epoch a correlation between the
``mass deficit''
(the mass needed to bring a flat inner density profile to a $r^{-2}$ cusp)
and the mass of the nuclear MBH, with a normalization and slope
comparable to the observed
relation (see Fig. \ref{fig4}). Models in which the mass displaced by the MBH binary
is replenished after every major galaxy merger appear instead to
underestimate the mass deficit observed in ``core" galaxies.  In \cite{vmh} a 
simple scheme was applied to hardening pairs, in which the ``loss cone" is constantly 
refilled and a constant density
core forms due to the ejection of stellar mass. The effect of
loss-cone depletion (the depletion of low-angular momentum stars that get close
enough to extract energy from a hard binary) is one of the major uncertainties in
computing the decay timescale, and makes it difficult to construct detailed
scenarios for coalescing black hole binaries.

\begin{figurehere}
\vspace{0.4cm}
\centerline{
\psfig{figure=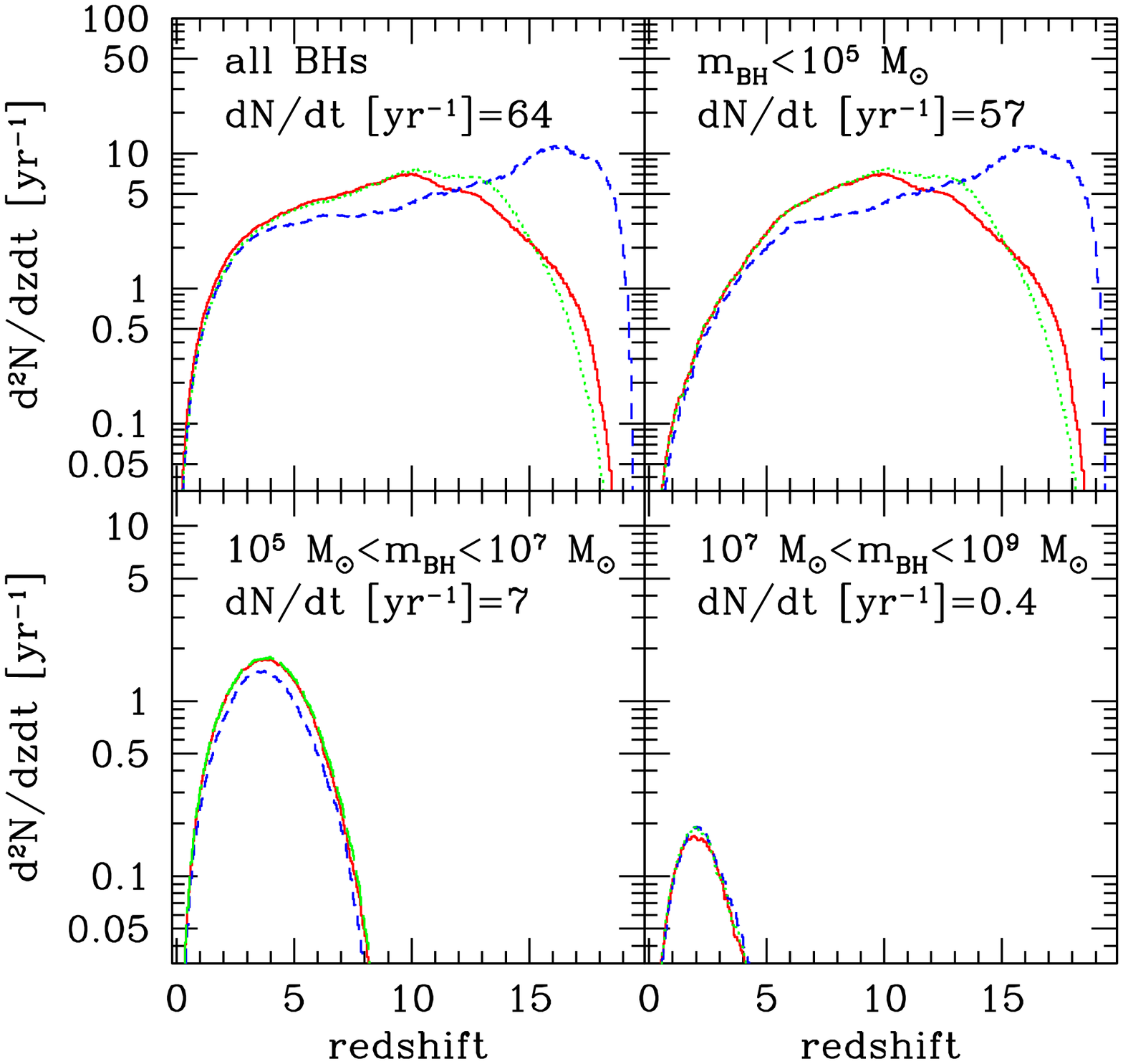,width=3.0in}}
\vspace{0.cm}
\caption{\footnotesize
Number of MBH binary coalescences observed per year at $z=0$, per unit redshift, in different
$m_{\rm BH}=M_1+M_2$ mass intervals. Each panel also lists the
integrated event rate, $dN/dt$, predicted by \cite{ses1}. The rates ({\it
solid lines}) are
compared to a case in which triple black hole interactions are switched off
({\it dotted lines}). Triple hole interactions increase the coalescence rate
at very high redshifts, while, for $10<z<15$, the rate is decreased because
of the reduced number of surviving binaries. {\it Dashed lines}: rates
computed assuming binary hardening is instantaneous, i.e. MBHs coalesce
after a dynamical friction timescale.
}
\label{fig5}
\vspace{0.4cm}
\end{figurehere}

\section{Gravitational radiation from inspiraling MBH binaries}

MBH binaries, with masses in the range $10^{3}-10^{7}\,\msun$, are one of the 
primary target for {\it LISA} \cite{h94}\cite{wl}\cite{ses1}. Interferometers operate 
as all-sky monitors, and the data streams collect
the contributions from a large number of sources belonging to
different cosmic populations. To optimize
the subtraction of resolved sources from the data stream, it is important to
have a detailed description of the expected rate, duration, amplitude,
and waveforms of events. Figure \ref{fig5} shows the
number of MBH binary coalescences per unit redshift per unit {\it observed} year predicted 
by \cite{ses1}, using a detailed model of MBH binaries dynamics. The observed event rate 
is obtained by dividing the rate per unit proper time by the $(1+z)$ cosmological time
dilation factor. Each panel shows the rate for
different $m_{\rm BH}=M_1+M_2$ mass intervals, and lists the integrated
event rate, $dN/dt$, across the entire sky. The number of events per observed
year per unit redshift peaks at $z=2$ for $10^7<m_{\rm BH}<10^9\,\msun$, at
$z=3-4$ for $10^5<m_{\rm BH}<10^7\,\msun$, and at $z=10$ for
$m_{\rm BH}<10^5\,\msun$, i.e. the lower the black hole mass, the higher the
peak redshift. Beyond the peak, the event rate decreases steeply
with cosmic time. 

In the stationary case, i.e., assuming no orbital decay, the GW emission
spectrum of a MBH pair in a circular orbit of radius $a$ is a delta function
at rest-frame frequency $f_r=\omega/\pi$, where $\omega=\sqrt{G(M_1+M_2)/a^3}$
is the Keplerian angular frequency of the binary. Orbital decay due to GW emission 
results in a shift of the emitted frequency to increasingly larger values as 
the binary evolution proceeds. Typically, the timescale for frequency shift is 
long compared to the wave period, and short compared to the duration of the
observation. Only close to the innermost stable circular orbit (ISCO), the GW
frequency changes at a rate comparable to the frequency itself. 
The rest-frame energy flux (energy per unit
area per unit time) associated to the GW is
\begin{equation}
\frac{dE}{dAdt}=\frac{\pi}{4}\frac{c^3}{G}f_r^2 h^2, 
\label{eqenergyflux}
\end{equation}
where the strain amplitude (sky-and-polarisation averaged) at comoving distance $r(z)$ 
is 
\begin{equation}
h\,=\,\frac{8\pi^{2/3}}{10^{1/2}}\,\frac{G^{5/3}\mathcal{M}^{5/3}}{c^4r(z)}
\,f_r^{2/3}, \label{eqstrain}
\end{equation}
$\mathcal{M}$ is the ``chirp mass'' of the binary, and all the
other symbols have their standard meaning. The strain is averaged
over a wave period. The important quantity to consider is the number 
of cycles $n$ spent in a frequency interval $\Delta f \simeq f$
around a given frequency $f$. In general, $n=f^2/\dot f\propto f^{-5/3}$. 
For a periodic signal at frequency $f$ lasting
for a time interval longer than the observation time $\tau$, we
have simply $n=f\tau$. The characteristic strain in an observation of (observed)
duration $\tau$ is then
\begin{equation}
h_c=h\sqrt{n} \propto f_r^{-1/6}, \qquad n<f\tau,
\label{eq1h_c}
\end{equation}
and
\begin{equation}
h_c=h\sqrt{f\tau}\propto f_r^{7/6}, \qquad n>f\tau,
\label{eq2h_c}
\end{equation}
where $f=f_r/(1+z)$ is the observed frequency.
In Figure \ref{fig6} $h_c$ is plotted for different MBH binaries
at different redshifts, compared to the {\it LISA} $h_{\rm rms}$
multiplied by a factor of 5, assuming a 3-year observation.

\begin{figurehere}
\vspace{0.4cm}
\centerline{
\psfig{figure=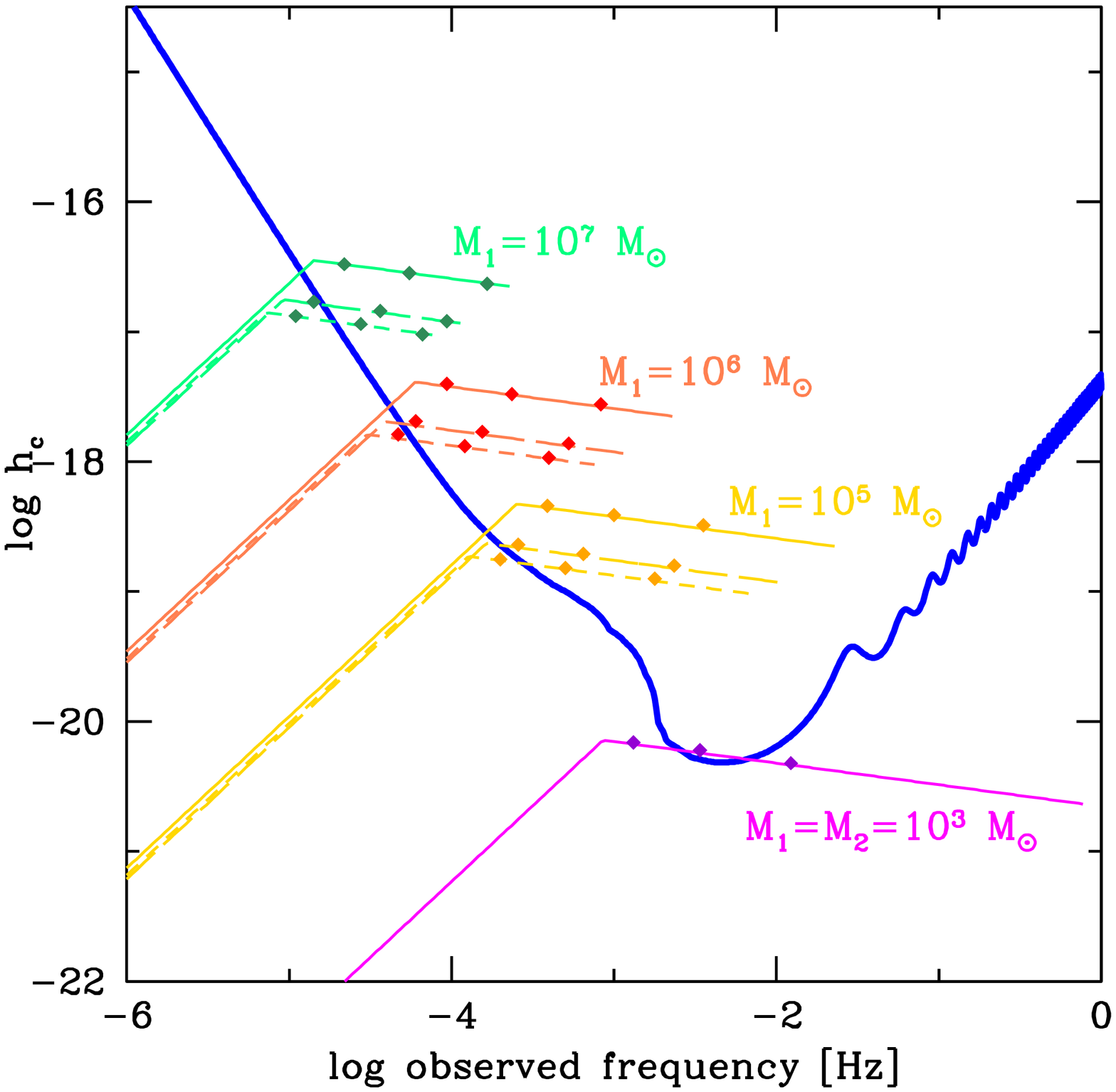,width=2.8in}}
\vspace{0.cm}
\caption{\footnotesize Characteristic strain $h_c$ for
MBH binaries with different masses and redshifts. From
top to bottom, the first three curves refer to systems with
$\log(M_1/\msun)=7,6,5$, respectively, and
$M_2=0.1 M_1$. The solid, long-dashed, and short-dashed lines
assumes the binary at $z=1,3,5$, respectively. A 3-year
observation is considered. The lowest solid curve assumes an equal mass
binary $M_1=M_2=10^3\,\msun$ at $z=7$. The small diamonds on
each curve mark, from left to right, the observed frequency at 1
year, 1 month
and 1 day before coalescence. The thick curve is {\it LISA} $5h_{\rm
rms}$, approximatively the threshold for detection
with $S/N \geq 5$. (From \cite{ses2}).
}
\label{fig6}
\vspace{0.4cm}
\end{figurehere}

At frequencies higher than the ``knee'', the time spent around a given
frequency is less than 3 years, and $h_c\propto f^{-1/6}$. The
signal shifts toward higher frequency during the
observation, and reaches the ISCO and the coalescence
phase in most cases. The lowest curve represents a low mass, high redshift equal
mass binary. In terms of their detectability by {\it LISA},
they represent a somewhat different class of events. Contrary to
the case of more massive binaries present at lower $z$, the final
coalescence phase of light binaries lies at too high frequecies,
well below the {\it LISA} threshold.
For frequencies much below the knee, the characteristic strain is
proportional to $f^{7/6}$, as the timescale for frequency shift is
longer than 3 years.
The signal amplitude is then limited by the observation time, not
by the intrinsic properties of the source. The source will be
observed as a ``stationary source", a quasi-monochromatic wave for
the whole duration of the observation. An increase in the observation
time will result in a shift of the knee toward lower frequencies. The
time needed for the sources to reach the ISCO starting from the
knee frequency is, approximatively, the observing time.

As recently shown by \cite{ses2}, the GW signal from MBH binaries will be resolved 
(assuming a 3-year {\it LISA} observation) into $\sim 100$ discrete events,
40 of which will be observed above threshold until coalescence. These ``merging events" involve
relatively massive binaries, $M\sim 10^5 \msun$, in the redshift
range $2 \lta z \lta 6$. The remaining $\sim 60$ events come
from higher redshift, less massive binaries ($M \sim 5\times 10^3 \msun$
at $z \gta 6$) and, although their $S/N$ integrated over the duration of
the observation can be substantial, the final coalescence phase is at too
high frequency to be directly observable by {\it LISA}. The total number
of detected events accounts for a fraction $\gta 90$\% of
all coalescences  of massive black hole binaries at $z\lta 5$.
The residual confusion noise from unresolved
massive black hole binaries is expected to be at least an
order of magnitude below the estimated stochastic {\it LISA} noise.

\section{MBH spins}

The spin of a MBH is determined by the competition between a number of
physical processes.  Black holes forming from the gravitational
collapse of very massive stars endowed with rotation will in general
be born with non-zero spin \cite{fwh}. An
initially non-rotating hole that increases its mass by (say) 50\% by
swallowing material from an accretion disk may be spun up to $a/m_{\rm
BH}=0.84$ \cite{bar70}. While the coalescence of two non-spinning
black holes of comparable mass will immediately drive the spin
parameter of the merged hole to $a/m_{\rm BH}\gta 0.8$ \cite{gam}, the capture 
of smaller companions in randomly-oriented orbits may spin down a 
Kerr hole instead \cite{hb}. 

\begin{figurehere}
\vspace{0.4cm}
\centerline{
\psfig{figure=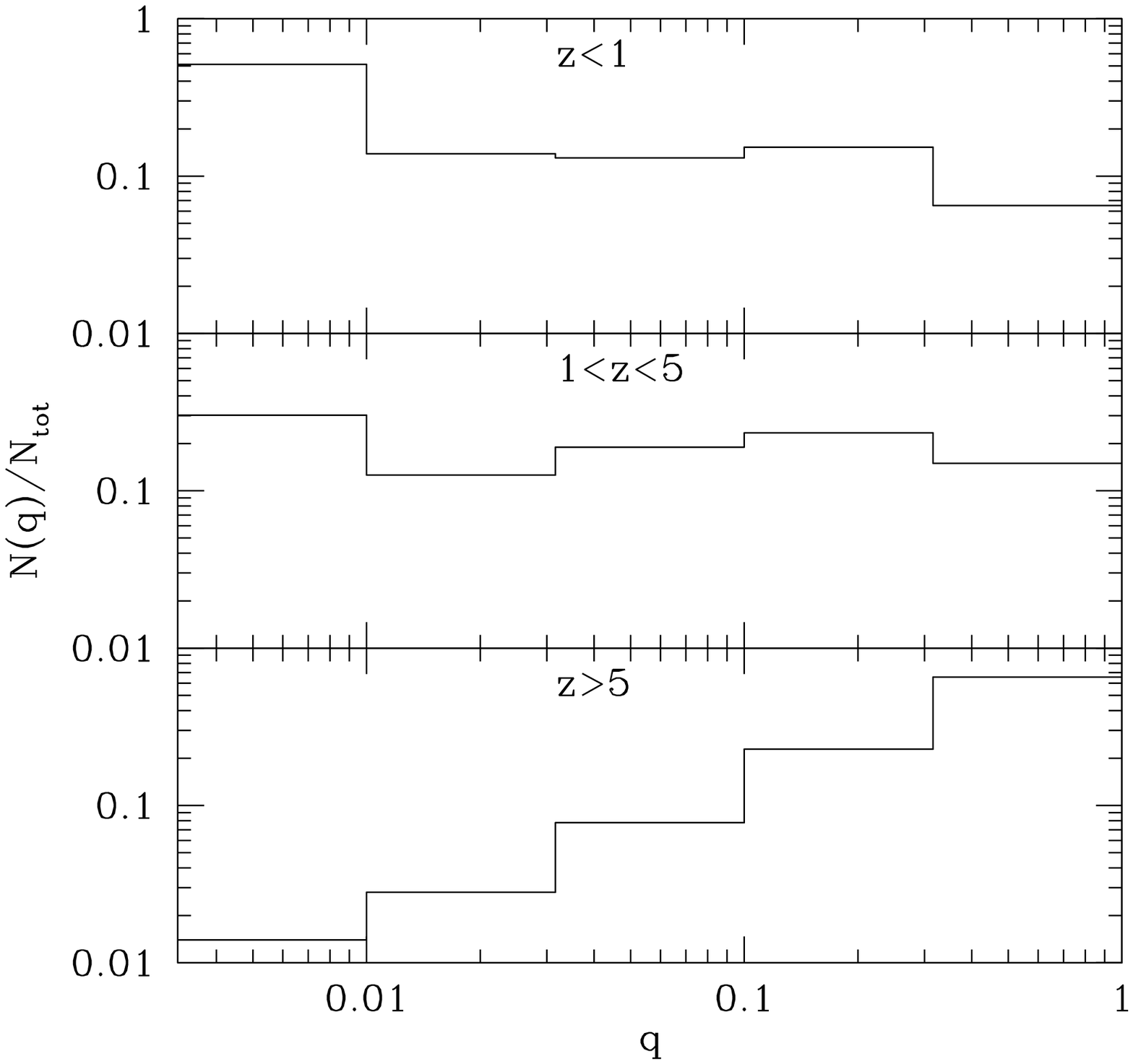,width=2.8in}}
\vspace{0.cm}
\caption{\footnotesize Normalized distribution of mass ratios, $q=M_2/M_1$, of 
coalescing MBH binaries at three different epochs. Note that at low
redshift MBHs typically capture much smaller companions.}
\label{fig7}
\vspace{0.4cm}
\end{figurehere}

In \cite{volo4} we have made a first attempt at estimating 
the distribution of MBH spins and its evolution with cosmic time in the context of
hierarchical structure formation theories, following the combined effects of black
hole-black hole coalescences and accretion from a gaseous disk on the
magnitude and orientation of MBH spins. Here I will briefly summarize our findings. 
Binary coalescences appear to cause no significant systematic
spin-up or spin-down of MBHs: because of the relatively flat
distribution of MBH binary mass ratios in hierarchical models (shown in Fig. \ref{fig7}), 
the holes random-walk around the spin parameter they are endowed with at birth, and 
the spin distribution retains significant memory of the initial rotation of ``seed'' holes.

It is accretion, not binary coalescences, that dominates the spin
evolution of MBHs (Fig. \ref{fig8}). Accretion can lead to efficient spin-up of MBHs
even if the angular momentum of the inflowing material varies in
time. This is because, for a thin accretion disk, the hole is aligned
with the outer disk on a timescale that is much shorter than the
Salpeter time \cite{np98}, leading
to accretion via prograde equatorial orbits.  As a result, most of the
mass accreted by the hole acts to spin it up, even if the orientation
of the spin axis changes in time.  For a geometrically thick disk,
alignment of the hole with the outer disk is much less efficient,
occurring on a timescale comparable to the Salpeter time.  Even in this case most 
holes will be rotating rapidly. This is because, in any model in which MBH growth
is triggered by major mergers, every accretion episode must typically
increase a hole's mass by about one e-folding to account for the local
MBH mass density and the $m_{\rm BH}-\sigma_*$ relation.  Most
individual accretion episodes thus produce rapidly-rotating holes
independent of the initial spin.

\begin{figurehere}
\vspace{0.4cm}
\centerline{
\psfig{figure=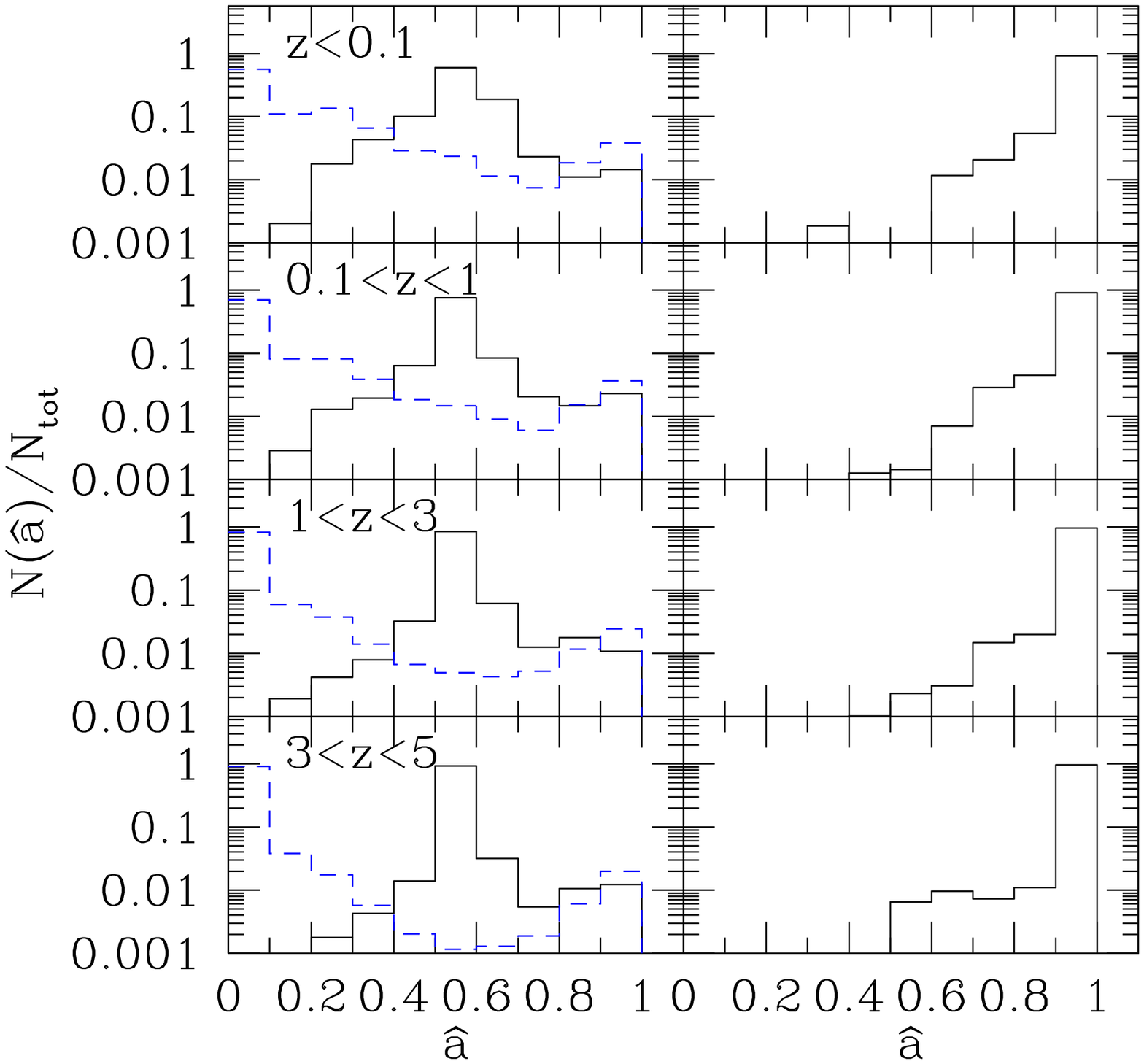,width=2.8in}}
\vspace{0.cm}
\caption{\footnotesize Distribution of MBH spins in different redshift intervals.
{\it Left panel:} effect of black hole binary coalescences only.
{\it Solid histogram:} seed holes are born with $\hat a\equiv a/m_{\rm BH}=0.6$.
{\it Dashed histogram:} seed holes are born non-spinning.
{\it Right panel:} spin distribution from binary coalescences and gas accretion.
Seed holes are born with $\hat a=0.6$, and are efficiently spun up
by accretion via a thin disk.
}
\label{fig8}
\vspace{0.4cm}
\end{figurehere}

Under the combined effects of accretion and binary coalescences, we
find that the spin distribution is heavily skewed towards
fast-rotating Kerr holes, is already in place at early epochs, and
does not change significantly below redshift 5.  As shown in Figure
\ref{fig9}, about 70\% of all MBHs are maximally rotating and have
mass-to-energy conversion efficiencies approaching 30\%.
Note that if the equilibrium spin attained by accreting MBHs is lower than
the value of $\hat a \equiv a/m_{\rm BH}= 0.998$ used here, as in the thick disk MHD
simulations of \cite{gam}, where $\hat a\approx 0.93$, then the
accretion efficiency will be lower as well, $\approx 17\%$.

\begin{figurehere}
\vspace{0.0cm}
\centerline{
\psfig{figure=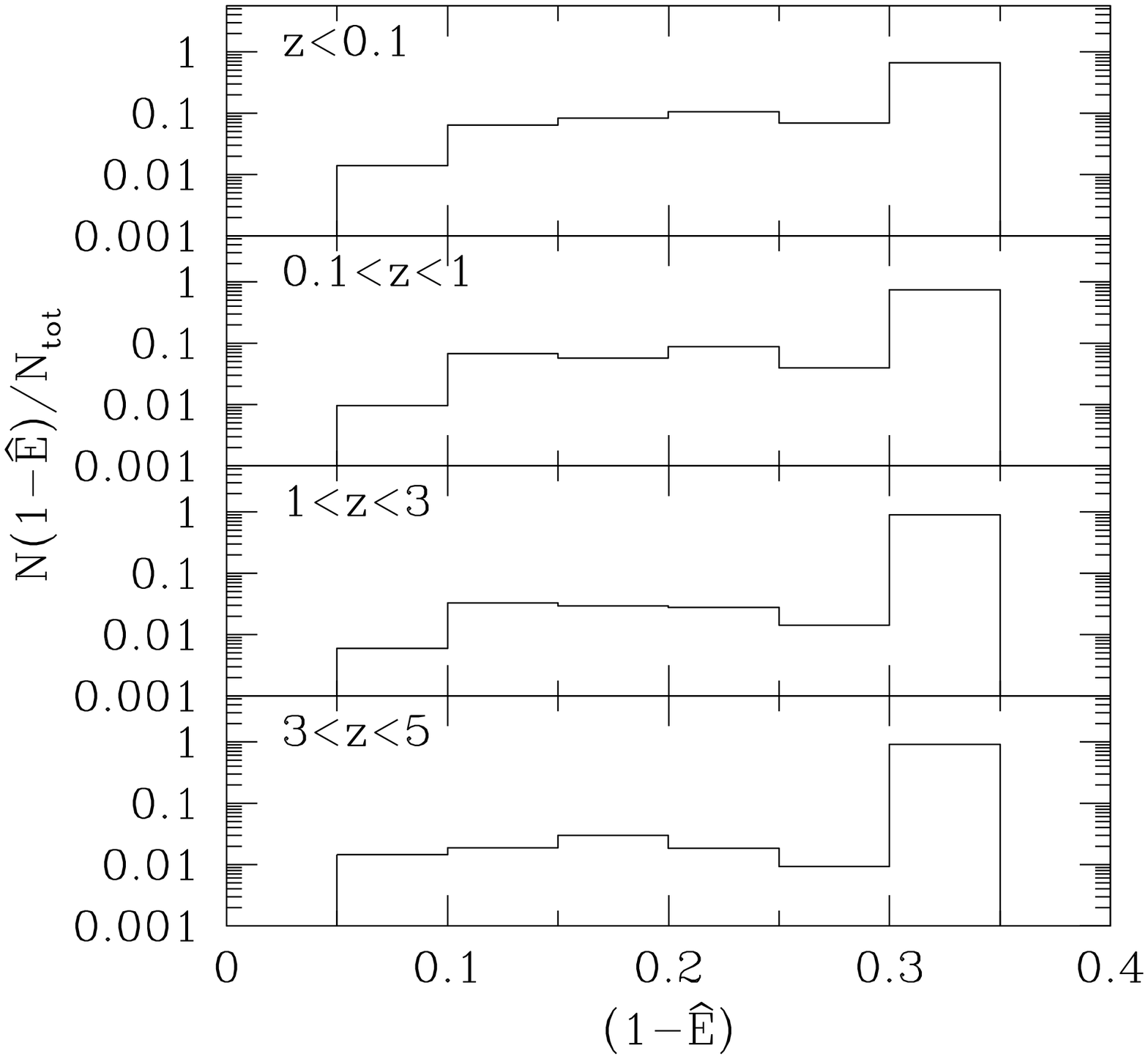,width=2.8in}}
\vspace{0.cm}
\caption{\footnotesize Distribution of accretion efficiencies,
$\epsilon\equiv 1-\hat E$, in
different redshift intervals (assuming that the energy radiated is the
binding energy at the ISCO). The spin distribution from binary
coalescences and gas accretion has been calculated assuming the holes
accrete via a thin disk on prograde equatorial orbits.}
\label{fig9}
\vspace{0.4cm}
\end{figurehere}

Even in the conservative case where accretion is via a geometrically thick disk
(and hence the spin/disk alignment is relatively inefficient) and the
initial orientation between the hole's spin and the disk rotation axis
is assumed to be random, we find that most MBHs rotate rapidly with
spin parameters $\hat a >0.8$ and accretion efficiencies
$\epsilon>$12\%.  As recently shown by \cite{yt}\cite{mar04}, 
a direct comparison between the local MBH mass density
and the mass density accreted by luminous quasars shows that quasars
have a mass-to-energy conversion efficiency $\epsilon\gta 0.1$ (a
simple and elegant argument originally provided by \cite{sol}).  This
high average accretion efficiency may suggest rapidly rotating Kerr
holes, in agreement with our findings. Since most holes rotate rapidly at all
epochs, our results suggest that spin is not a necessary and
sufficient condition for producing a radio-loud quasar.

\bigskip\bigskip 
I would like to thank my numerous collaborators on this subject: F. 
Haardt, M. Kuhlen, P. Oh, E. Quataert, M. Rees, A. Sesana, and M. Volonteri. 
This manuscript was written while the author was enjoying the hospitality of 
the Kavli Institute for Theoretical Physics. Support for this work was provided 
by NASA grant NNG04GK85G, and by NSF grants AST-0205738 and PHY99-07949.

%

\end{document}